\documentclass{optica-article}
\journal{oe}

\usepackage{lineno}

\begin{document}

\title{Photothermal effect in macroscopic optomechanical systems with an intracavity nonlinear optical crystal}

\author{Sotatsu Otabe,\authormark{1,*} Kentaro Komori,\authormark{2} Ken-ichi Harada,\authormark{1} Kaido Suzuki,\authormark{1} Yuta Michimura,\authormark{3,4} and Kentaro Somiya\authormark{1}}

\address{\authormark{1}Department of Physics, Tokyo Institute of Technology, Meguro, Tokyo 152-8550, Japan\\
\authormark{2}Department of Physics, University of Tokyo, Bunkyo, Tokyo 113-0033, Japan\\
\authormark{3}LIGO Laboratory, California Institute of Technology, Pasadena, California 91125, USA\\
\authormark{4}Research Center for the Early Universe (RESCEU), Graduate School of Science, University of Tokyo, Tokyo 113-0033, Bunkyo, Japan}
\email{\authormark{*}otabe@gw.phys.titech.ac.jp}

\begin{abstract}
	Intracavity squeezing is a promising technique that may improve the sensitivity of gravitational wave detectors and cool optomechanical oscillators to the ground state. However, the photothermal effect may modify the occurrence of optomechanical coupling due to the presence of a nonlinear optical crystal in an optical cavity. We propose a novel method to predict the influence of the photothermal effect by measuring the susceptibility of the optomechanical oscillator and identifying the net optical spring constant and photothermal absorption rate. Using this method, we succeeded in precisely estimating parameters related to even minor photothermal effects, which could not be measured using a previously developed method.
\end{abstract}

\section{Introduction}
The application of cavity optomechanics~\cite{RevModPhys.86.1391} is a valuable technique used to examine the quantum nature of macroscopic objects. By coupling a mechanical oscillator to an optical cavity using a strong optical field, a wide range of test masses have been evaluated. Sideband cooling~\cite{Arcizet2006,Gigan2006,PhysRevLett.97.243905} is a powerful and well-established technique for cooling a mechanical oscillator to its quantum ground state by using optomechanical coupling~\cite{Teufel2011,Chan2011,Verhagen2012,PhysRevA.92.061801,PhysRevLett.116.063601,PhysRevLett.124.173601}. Because the stabilized laser is at thermal equilibrium with the very low-temperature bath, it is possible to significantly decrease the ambient temperature of optomechanical systems by generating a low noise-damping source with light~\cite{Gigan2006}.\par
Optomechanical coupling is also a promising technique for interferometric gravitational wave detectors~\cite{Harry_2010, Acernese_2014, Willke_2006, Somiya_2012}. While the imaginary component of the optical spring plays a vital role in sideband cooling, the real component of the optical spring improves the sensitivity~\cite{Buonanno_2001,PhysRevD.64.042006,PhysRevD.65.042001}. When the optical cavity in the interferometer is slightly detuned from resonance, a fraction of the gravitational wave signal field couples to the laser field to generate a radiation pressure force on the test mass. The optical spring then enhances the gravitational wave signal, and the signal-to-noise ratio against quantum noise is improved around the resonant frequency~\cite{PhysRevD.65.042001,Chen_2013}. \par
Although optical springs are used in various applications, a simultaneous increase in the real and imaginary components of the oscillator is not possible because the real and imaginary components of the complex optical spring constant have opposite signs~\cite{RevModPhys.86.1391,Arcizet2006,PhysRevD.65.042001}. An optical spring alone is always unstable; therefore, the addition of a mechanical spring or a supplementary control mechanism is essential. Moreover, the most significant problem associated with the effective implementation of optical springs is that the intracavity power limits the impact of the optical spring. It can be challenging to generate an optical spring with firm damping or a strong restoring force without compromising the performance of the interferometer because increasing the cavity finesse narrows the bandwidth of the cavity, while increasing the input power induces thermal lensing or other harmful effects~\cite{Teufel2011,PhysRevLett.116.063601,Harry_2010,Willke_2006}.\par
To solve this problem, the implementation of a technique called intracavity squeezing was originally proposed to generate a stiff optical spring~\cite{SOMIYA2016521,KOROBKO20182238}. This technique has been studied as a method for widening the bandwidth of gravitational wave detectors~\cite{PhysRevLett.95.193001,PhysRevLett.118.143601,Korobko2019} and can reinforce only the signal response of the cavity without increasing the intracavity power. Intracavity squeezing method can also effectively cool down a macroscopic mechanical oscillator to its quantum ground state~\cite{PhysRevA.79.013821,Asjad:19,https://doi.org/10.1002/lpor.201900120}. It not only enhances the occurrence of optical damping but also induces quantum noise interference for all dissipation ports, eliminating quantum backaction even in the unresolved sideband regime.\par
Intracavity squeezing method inevitably introduces a photothermal effect in nonlinear optical crystals. Various studies have been conducted on the influence of photothermal effects on laser interferometers. Such photothermal effects acting in the cavity can enable the self-locking of the cavity~\cite{Carmon:04,Altin2017} or conversely induce instabilities~\cite{PhysRevA.95.013826,Ma2020}. The force exerted by the photothermal effect on the test mass is also referred to as the bolometric force~\cite{Pinard_2008,PhysRevA.83.033809,RESTREPO2011860,PhysRevA.86.043803}, which has been used for optomechanical cooling~\cite{Gigan2006,Metzger2004,PhysRevB.78.035309,PhysRevLett.101.133904}. Another interesting recent application of the photothermal effect is {\it photothermally induced transparency}~\cite{doi:10.1126/sciadv.aax8256,Clementi2021}, which is a result of the photothermal effect modifying the effective cavity length. Analogous to the well-known phenomena of electromagnetically induced transparency~\cite{PhysRevLett.66.2593,RevModPhys.77.633} and optomechanically induced transparency~\cite{PhysRevA.81.041803,doi:10.1126/science.1195596,Safavi-Naeini2011}, a cavity is realized with an extremely narrow linewidth through the coupling of the optical cavity and an intracavity object. The same research group that developed photothermally induced transparency has also demonstrated that the photothermal effect changes the optical response of the cavity~\cite{Ma:21}.\par
In this study, we investigate the influence of the photothermal effect in an optomechanical system with a nonlinear optical crystal. The photothermal effect modifies the complex optical spring constant~\cite{PhysRevD.92.062003,Altin2017,Qin:22} as well as the optical response of the cavity. Specifically, the displacement of the test mass does not match the effective cavity length modified by the photothermal effect, resulting in partial exchange of the real and imaginary components of the optical spring, which may significantly distort the susceptibility of the optomechanical oscillator. The photothermal effect of a nonlinear optical crystal is much more significant than that resulting from cavity optics, which was reported in a previous study~\cite{doi:10.1126/sciadv.aax8256,Ma:21}; indicating the importance of the problem in the intracavity squeezing systems. To evaluate the effect of intracavity squeezing, it is essential to accurately predict the photothermal effect acting on a nonlinear optical crystal.\par
The nature of the photothermal effect is determined by the characteristic frequencies of thermal absorption and relaxation in the examined system. These parameters can be obtained via optical response measurements of a cavity~\cite{Ma:21}. However, when thermal relaxation occurred sufficiently faster than thermal absorption, the photothermal effect quickly reached an equilibrium and the photothermal parameters cannot be easily measured. Therefore, we propose a new method to simultaneously estimate the photothermal absorption rate and optical spring constant by measuring the susceptibility of an optomechanical oscillator over a wide frequency bandwidth. Our method accurately estimates the magnitude of even a minor photothermal effect because a small amount of thermal absorption results in non-negligible optical damping and serves to predict the photothermal effect in various optical systems with an absorptive crystal.

\section{Principle}\label{sec:PT-THEORY}

\begin{figure}[htbp]
	\centering
	\includegraphics[width=\hsize]{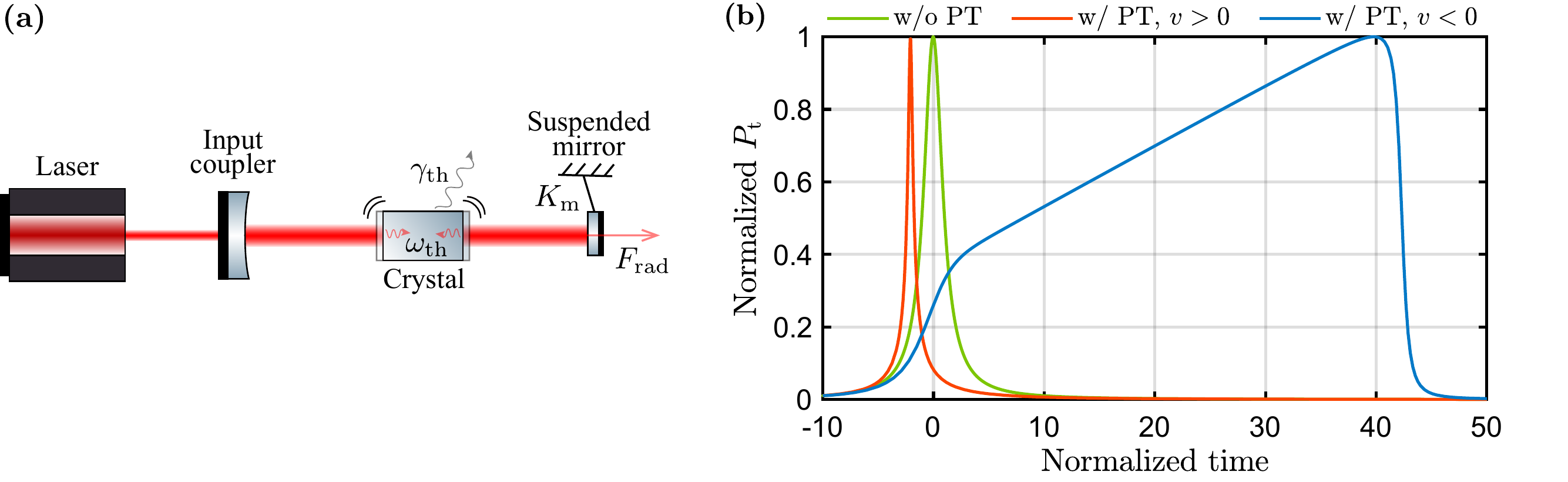}
	\caption{\textbf{(a)} Schematic of an optical cavity with an absorptive crystal. The characteristic frequencies of photothermal absorption and relaxation are $\omega_\mathrm{th}$ and $\gamma_\mathrm{th}$, respectively. One mirror constituting the cavity is suspended by a mechanical spring with a complex spring constant $K_\mathrm{m}$ and receives a radiation pressure force $F_\mathrm{rad}$ from the intracavity light field. \textbf{(b)} Simulation results of the transmitted power from the cavity. The displacement of the mirror $x_\mathrm{act}$ varies at a constant velocity $v$. The green line shows the response without the photothermal effect and exhibits a Lorentzian curve. The red and blue lines show the responses with the photothermal effect when the velocity is positive and negative, respectively. The vertical axis represents transmitted power, normalized to be one at the resonance. The horizontal axis is the normalized time such that the half-width of the spectrum excluding the photothermal effect is 1.}
	\label{fig:PT-outline}
\end{figure}

In this section, we discuss the frequency response of an optomechanical system containing a crystal that causes a photothermal effect, as shown in Fig.\,\ref{fig:PT-outline}(a). After reviewing the fundamental equations for the photothermal effect, we derive the optical response of the cavity and susceptibility of the optomechanical oscillator with the photothermal effect.
	
\subsection{Fundamental equations of the photothermal effect}
Regardless of the existence of an absorptive crystal, the intracavity power $P$ can be written as:
\begin{equation}
	P=\dfrac{2\mathcal{F}}{\pi}\dfrac{1}{1+\xi^2}P_0,\label{eq:PT-intracavitypower}
\end{equation}
where $\mathcal{F}$ is the cavity finesse and $P_0$ is the carrier power incident on the cavity. We assume that the transmissivity of the input coupler is sufficiently larger than other internal losses and the cavity is sufficiently close to the resonant state. $\xi$ is the normalized cavity detuning, which is proportional to the effective cavity length change $x=x_\mathrm{act}+x_\mathrm{th}$ where $x_\mathrm{act}$ is the actual displacement of the test mass and $x_\mathrm{th}$ is the effective change in the cavity optical path length owing to the photothermal effect:
\begin{equation}
	\xi=\dfrac{2\mathcal{F}\omega_0}{\pi c}x.\label{eq:PT-normalizeddetune}
\end{equation}
Here, $\omega_0$ is the angular frequency of the carrier field and $c$ is the speed of light. The photothermal displacement $x_\mathrm{th}$ is given by $x_\mathrm{th}=\alpha L'(T-T_0)$, with $\alpha$ being the thermal expansion coefficient including the thermo-optic effect (refractive index change with temperature), $L'$ being the crystal length, $T$ being the temperature of the part of the crystal contributing to the photothermal effect (hereafter referred to as the crystal), and $T_0$ being the surrounding temperature. Consequently, over time $t$, $\partial x_\mathrm{th}/\partial t$ is proportional to $\partial T/\partial t$ and thus to the net heat obtained by the crystal: 
\begin{equation}
	\dfrac{\partial x_\mathrm{th}}{\partial t}=\alpha L'\dfrac{\partial T}{\partial t}=\dfrac{\alpha L'}{C}(w-q).\label{eq:PT-thermalflow}
\end{equation}
Here, $C$ is the heat capacity of the crystal and $w$ and $q$ are the time rate of heat flow into and out of the crystal, respectively. The absorption of the carrier light causes the constant thermal inflow into the crystal, which can be written as $w=\alpha'L'P$ with the absorption coefficient $\alpha'$. The heat outflow can be divided into two parts: one due to the heat conduction or heat transfer and the other due to thermal radiation. However, if the difference between $T$ and $T_0$ is small, we can neglect the latter component and the heat outflow can be written as $q=(T-T_0)/k$ where $k$ is the thermal resistance.
	
\subsection{Optical response of the cavity with the photothermal effect}\label{sec:PT-TFofcavity}
First, we consider the optical response of the cavity. In the optical system shown in Fig.\,\ref{fig:PT-outline}(a), the cavity spectrum is not solely Lorentzian because the effective cavity length changes owing to the photothermal effect. Figure\,\ref{fig:PT-outline}(b) shows the transmitted power from the cavity when the mirror moves at a constant velocity $v=\partial x_\mathrm{act}/\partial t$. If the mirror moves in the same direction as the photothermal displacement to increase thermal absorption, the cavity reaches the resonant point faster than it would in the absence of the photothermal effect, and the linewidth of the spectrum narrows. If the mirror moves in the opposite direction, the photothermal effect cancels the mirror motion until the cavity reaches the resonant point, and the linewidth of the spectrum shows an enormously broadened response~\cite{Carmon:04,Altin2017,PhysRevA.95.013826,doi:10.1126/sciadv.aax8256,Ma:21,Qin:22}.\par
The photothermal effect also modifies the frequency response of the cavity with the reciprocal of the photothermal absorption and relaxation time scale as the characteristic frequency~\cite{Ma:21}, even in a frequency band that is sufficiently lower than the cavity decay rate. We derive the optical response of the cavity from the differential equation formed between the cavity detuning and displacement of the test mass. From Eqs.\,(\ref{eq:PT-intracavitypower}) to (\ref{eq:PT-thermalflow}), the following equation can be derived:
\begin{equation}
	\dfrac{\partial\xi}{\partial t}=-\dfrac{1}{kC}\xi+\dfrac{4\mathcal{F}^2\omega_0\alpha\alpha'L'^2P_0}{\pi^2 c C}\dfrac{1}{1+\xi^2}+\dfrac{2\mathcal{F}\omega_0}{\pi c}\left(\dfrac{1}{kC}x_\mathrm{act}+\dfrac{\partial x_\mathrm{act}}{\partial t}
	\right).\label{eq:PT-deltapart}
\end{equation}
The second term on the right-hand side is nonlinear in $\xi$ but can be linearized by splitting it into the stationary term $\xi_0$ and the relatively small fluctuating term $\delta\xi(t)$, i.e., $\xi(t)=\xi_0+\delta\xi(t)$. Only the first-order term of $\delta\xi(t)$ will be considered. We also perform the same operations to $x_\mathrm{act}$ and $x_\mathrm{th}$ to obtain $x_\mathrm{act}(t)=\bar{x}_\mathrm{act}+\delta x_\mathrm{act}(t)$ and $x_\mathrm{th}(t )=\bar{x}_\mathrm{th}+\delta x_\mathrm{th}(t)$, respectively. We then define $\bar{x}=\bar{x}_\mathrm{act}+\bar{x}_\mathrm{th}$ and $\delta x(t)=\delta x_\mathrm{act}(t)+\delta x_\mathrm{th}(t)$. By performing a Fourier transform with the angular frequency $\Omega$, the optical response of the cavity $H_\mathrm{th}$ is derived as:
\begin{equation}
	H_\mathrm{th}=\dfrac{\delta x(\Omega)}{\delta x_\mathrm{act}(\Omega)}=\dfrac{\gamma_\mathrm{th}+i\Omega}{(\omega_\mathrm{th}+\gamma_\mathrm{th})+i\Omega},\label{eq:PT-cavityTF}
\end{equation}
with
\begin{equation}
    \omega_\mathrm{th}=\dfrac{8\mathcal{F}^2\omega_0\alpha\alpha'L'^2P_0}{\pi^2 c C}\frac{\xi_0}{(1+\xi_0^2)^2},\ \ \ \gamma_\mathrm{th}=\dfrac{1}{kC},\label{eq:PT-TFparameter}
\end{equation}
which are the photothermal absorption and relaxation rates, respectively. When the thermal absorption occurs faster than the thermal relaxation ($|\omega_\mathrm{th}|\gtrsim \gamma_\mathrm{th}$), the phase change owing to the photothermal effect is no longer negligible.\par
A qualitative explanation of the cavity behavior when $\alpha>0$ and $\xi_0>0$ is provided as follows. When the mirror moves at a frequency sufficiently higher than $\omega_\mathrm{th}$, the photothermal effect is not apparent because the signal reverses faster than the occurrence of the photothermal effect. When the mirror moves at a frequency comparable to $\omega_\mathrm{th}$, the effective cavity length accumulated by the photothermal effect is released as the signal changes, causing a phase lead in the optical response. When the mirror moves sufficiently slowly to cause photothermal relaxation, the phase does not change because the photothermal effect reaches equilibrium, but the gain is reduced because the photothermal effect cancels the effective cavity length change. It should also be noted that the sign of the pole of $H_\mathrm{th}$ could be reversed by changing the cavity detuning $\xi_0$.
	
\subsection{Susceptibility of the optomechanical oscillator with the photothermal effect}
The photothermal effect can contribute to optomechanical coupling in various ways. When mirror distortion due to thermal expansion is used to cool an optomechanical oscillator, the photothermal effect acts directly on the mechanical system as a bolometric force. Because the heating process owing to the photothermal effect defines the cooling limit of such a system, it is necessary to model the photothermal effect using time-delayed forces and then combine it with the semiclassical theory of photon absorption shot noise~\cite{PhysRevA.83.033809}.\par
In contrast, in the system shown in Fig.\,\ref{fig:PT-outline}(a), the photothermal effect affects only the effective cavity length, and no bolometric force acts on the test mass. When the radiation pressure force $F_\mathrm{rad}$ is proportional to the cavity length change $x$, denoted as $\delta F_\mathrm{rad}(\Omega)=-K_\mathrm{opt}(\Omega)\delta x(\Omega)$ wherein the proportionality factor $K_\mathrm{opt}(\Omega)$ is referred to as the complex optical spring constant, this relationship holds even in the presence of the photothermal effect. However, the displacement of the test mass does not match the effective cavity length change, as shown in Eq. \,(\ref{eq:PT-cavityTF}), and the complex optical spring constant for the test mass changes as $K_\mathrm{opt-th}=H_\mathrm{th}K_\mathrm{opt}$.\par
The effective susceptibility of the optomechanical oscillator $\chi_\mathrm{eff}$, which is the response from the external force $F_\mathrm{ext}$ that is applied to the test mass for the displacement of the test mass $x_\mathrm{act}$, can be written as:
\begin{equation}
	\chi_\mathrm{eff}=\dfrac{\delta x_\mathrm{act}(\Omega)}{\delta F_\mathrm{ext}(\Omega)}=\dfrac{1}{-m\Omega^2+K_\mathrm{m}+K_\mathrm{opt-th}},
\end{equation}
where $m$ is the effective mass of the suspended mirror and $K_\mathrm{m}$ is the complex mechanical spring constant. $K_\mathrm{m}$ can be written as $K_\mathrm{m}=k_\mathrm{m}+i\Gamma_\mathrm{m}\Omega$ using the mechanical spring constant $k_\mathrm{m}$ and mechanical damping constant $\Gamma_\mathrm{m}$. In addition, when the frequency band under consideration is sufficiently lower than the cavity decay rate $\gamma=\pi c/(2\mathcal{F}L)$ (where $L$ is the one-way length of the cavity), the complex optical spring constant can be written as $K_\mathrm{opt}=k_\mathrm{opt}+i\Gamma_\mathrm{opt}\Omega$ using the optical spring constant $k_\mathrm{opt}$ and optical damping constant $\Gamma_\mathrm{opt}$, in which the real component is:
\begin{equation}
	k_\mathrm{opt}\simeq\frac{16\mathcal{F}^2\omega_0P_0}{\pi^2 c^2}\frac{\xi_0}{(1+\xi_0^2)^2}.\label{eq:PT-kopt0}
\end{equation}
A mechanical spring and an optical spring are connected in parallel to the test mass, and the photothermal effect mixes the real and imaginary components of the effective complex optical spring constant. In particular, if the effect of $\Gamma_\mathrm{opt}$ on the complex spring constant is negligible, the photothermal effect converts the real component of the optical spring constant $k_\mathrm{opt}$ into the imaginary component as:
\begin{equation}
	K_\mathrm{opt-th}\simeq \frac{(\omega_\textrm{th}+\gamma_\textrm{th})\gamma_\textrm{th}+\Omega^2+i\omega_\textrm{th}\Omega}{(\omega_\textrm{th}+\gamma_\textrm{th})^2+\Omega^2}k_\mathrm{opt}.\label{eq:PT-Koptth}
\end{equation}\par
When $\alpha>0$ and $\xi_0>0$, the real and imaginary components of $K_\mathrm{opt-th}$ are positive, which indicates that a single carrier can generate a stable spring~\cite{PhysRevD.92.062003,Altin2017}. When the frequency band under consideration is sufficiently low ($\Omega\ll\omega_\mathrm{th},\gamma_\mathrm{th}$), the magnitude of the imaginary component is maximized when the photothermal absorption and relaxation rates are approximately equal ($\omega_\mathrm{th}\simeq\gamma_\mathrm{th}$). However, it should be noted that a slight photothermal effect could cause non-negligible optical damping.
	
\section{Experimental results}

\begin{figure}[htbp]
	\centering
	\includegraphics[width=0.8\hsize]{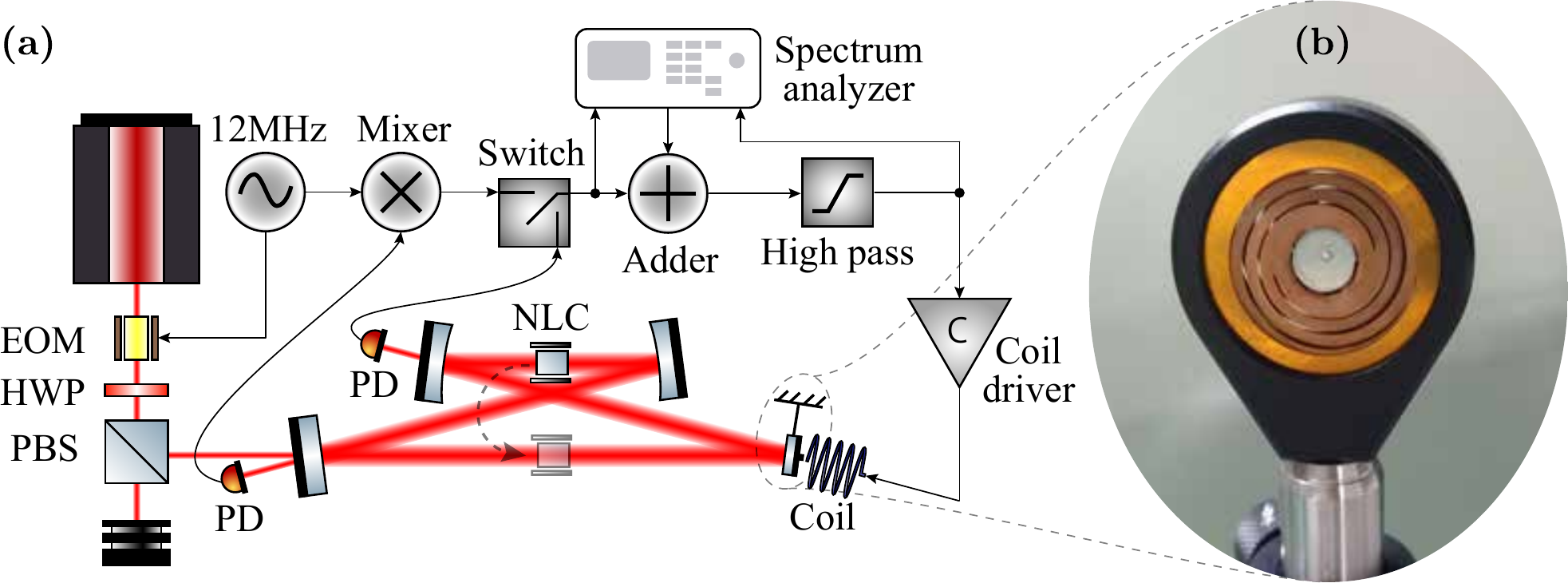}
	\caption{\textbf{(a)} Experimental setup. We used a 1064\,nm Nd:YAG laser as the light source and modulated the phase of incident light with an electro-optic modulator (EOM). We can adjust the intensity of the incident light using a half-wave plate (HWP) and a polarizing beam splitter (PBS). We measured the reflected and transmitted power of the cavity with photodetectors (PD) and for use as a control signal. We measured the transfer function of the optical system by adding the signal output from the spectrum analyzer using an adder and taking the signal ratio between the adder and driver inputs. We inserted a nonlinear optical crystal (NLC) through a 40-$\mu$m beam waist and re-installed them through a 300-$\mu$m beam waist when the thermal lens effect was seriously affected. \textbf{(b)} Suspended mirror. The mirror is $6.35$\,mm in diameter, weighs $280$\,mg, and is suspended by a double spiral spring. The suspension consists of two 0.5-mm thick beryllium copper (BeCu) plates, designed to exhibit a stiffened pitch and yaw by sandwiching the mirror between them from the front and back sides. We further sandwiched the suspension between brass rings to suppress higher-order mechanical resonances and glued neodymium magnets that were 1\,mm in diameter and 0.5-mm thick to the back of the mirror.}
	\label{fig:PT-setup}
\end{figure}
\subsection{Concept and setup of the experiment}
This experiment aimed to evaluate the influence of the photothermal effect on the intracavity squeezing system. The experimental setup is shown in Fig.\,\ref{fig:PT-setup}(a). We used a bowtie cavity with a nonlinear optical crystal, which was designed as an optical parametric oscillator cavity; however, one mirror was suspended by a double spiral spring, as shown in Fig.\,\ref{fig:PT-setup}(b). There were situations in which measurements needed to be performed without the influence of the optical spring. For these measurements, we replaced the suspended mirror by a piezoelectric element (PZT) and the high-pass filter by a low-pass filter. We used either the signal from the Pound-Drever-Hall technique~\cite{Drever1983} or the transmitted power as the error signal to control the cavity length, where the former was used for operating points with a slight cavity detuning $\xi_0$ and the latter for operating points with a large cavity detuning $\xi_0$.\par
This cavity has two mirrors with curvatures of 68.5\,mm. The reflectance of the mirrors is $94\pm1$\% for the input coupler, $99.95\pm0.02$\% for the curved mirror, and more than 99.8\% for the small mirror. The designed value of the circular length of the cavity is $2L=0.43$\,m. If we neglect intracavity losses caused by crystal and other factors, the cavity decay rate is approximately $\gamma\simeq1.1\times10^7$\,rad/s, and the finesse is approximately $\mathcal{F}\simeq 100$.\par
Two beam waists are present in the bowtie cavity. The beam radius of the waist between the two flat mirrors is 300\,$\mu$m and that between the two curved mirrors is 40\,$\mu$m. Unless otherwise noted, the nonlinear optical crystal is located at the latter waist. We used either periodically poled LiNbO$_3$ (PPLN) or periodically poled KTiOPO$_4$ (PPKTP) with a length of 10\,mm. In this experiment, the nonlinear optical crystal was set to an extreme phase-mismatch condition to evaluate only the photothermal effect. For example, the PPKTP crystal used in this experiment was phase-matched at approximately 35\,$^\circ$C, but we heated it to approximately 120\,$^\circ$C to collapse the phase-matching condition when measuring the transfer function. Therefore, nonlinear optical effects such as second-harmonic generation can be practically ignored. We measured the finesse in both crystals using weak incident light of approximately $5$\,mW and obtained $\mathcal{F}=100\pm10$.\par
Because the PPLN crystal has a relatively sizable photothermal absorption rate, it was useful in confirming the occurrence of the photothermal effect. Using this crystal, we report the measurement results of various phenomena caused by the photothermal effect in Sec.\,\ref{sec:PT-cavityresponsemeasurement}. Owing to the specificities of nonlinear optical crystals and the intracavity squeezing system, we have also succeeded in measuring interesting phenomena that have not yet been reported. Because PPKTP crystals have ideal characteristics for observing an optical spring with the photothermal effect, we report on the susceptibility measurement results obtained using this crystal in Sec.\,\ref{sec:PT-OSwPT}. In addition to confirming that the photothermal effect causes significant optical damping, we also successfully performed high-accuracy parameter estimation related to the optical spring and photothermal effect.
	
\subsection{Measurement of the photothermal effect}\label{sec:PT-cavityresponsemeasurement}

\begin{figure}[htbp]
	\centering
	\includegraphics[width=\hsize]{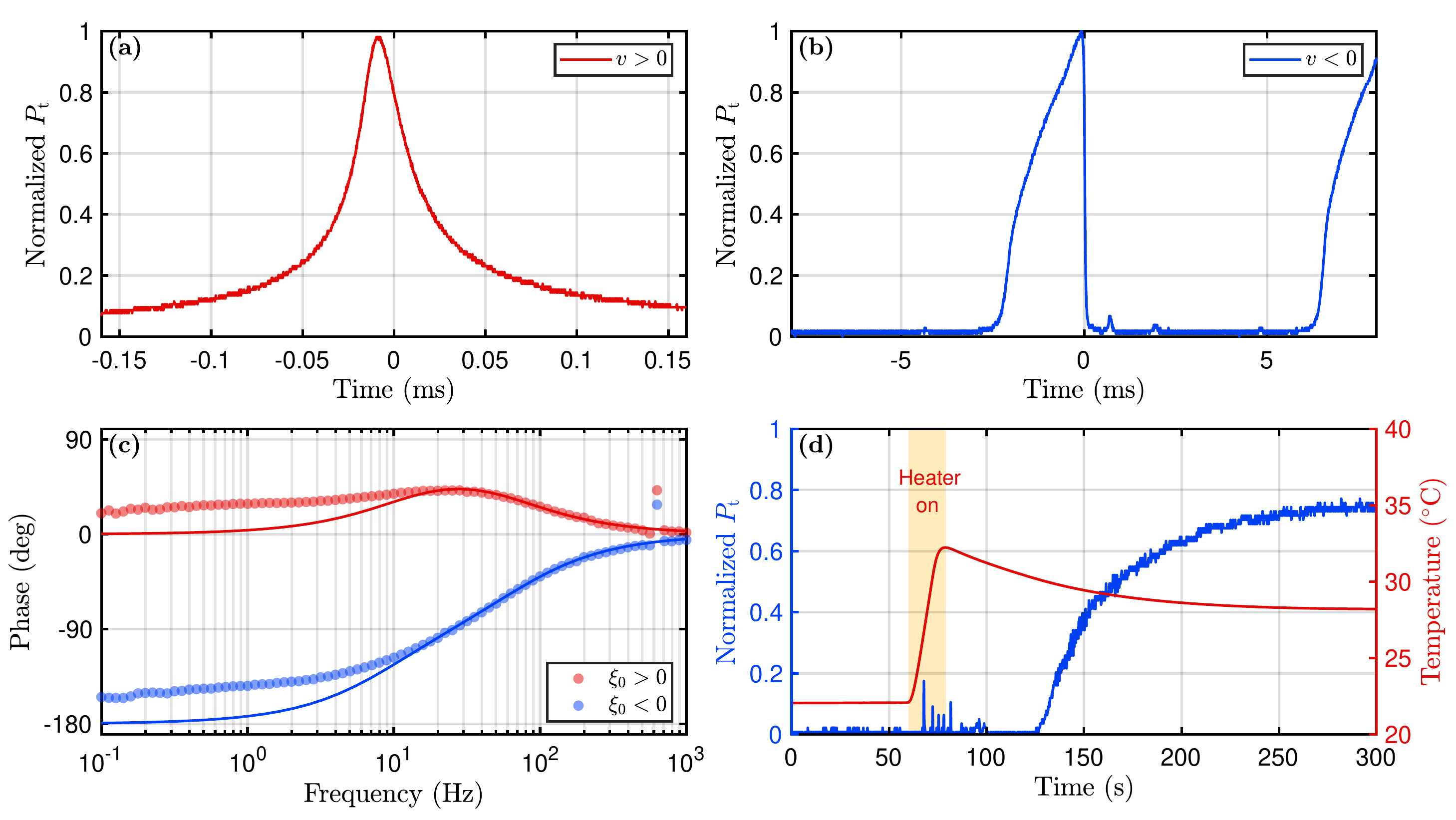}
	\caption{Measurement results of the photothermal effect. We performed these experiments using a PPLN crystal and piezoelectric actuator with an input power of 600\,mW. \textbf{(a)(b)} Transmitted power from the cavity scan. (a) corresponds to the case in which the speed of the cavity scan $v$ is positive, (b) corresponds to the case in which $v$ is negative, and we have adjusted the magnitude of each $v$ value to be almost identical. The vertical axis shows the transmitted power $P_\mathrm{t}$, normalized to be one at the resonant state. The horizontal axis shows time, but it should be noted that the scale differs by nearly two orders of magnitude in each case. \textbf{(c)} Phase response of the cavity with the photothermal effect. The red circles show the measurement results for the positive normalized cavity detuning $\xi_0$ ($\xi_0=1.00\pm0.02$), and the blue circles show the measurement results for the negative $\xi_0$ values ($\xi_0=-1.00\pm0.02$). The red and blue solid lines show the respective fitting results for the aforementioned data, in which we used only data in the band higher than 15\,Hz. \textbf{(d)} Self-locking of the cavity. The rightmost vertical axis and solid red line show the surface temperature of the crystal, while the leftmost vertical axis and solid blue line show the normalized transmitted power. We warmed the crystal at a constant heat over a time of $60\sim79$\,s, as indicated in yellow.}
	\label{fig:PT-spectrumetc}
\end{figure}

If the photothermal absorption rate $\omega_\mathrm{th}$ is sufficiently greater than the photothermal relaxation rate $\gamma_\mathrm{th}$, the photothermal effect can be easily confirmed in several ways. Here, we measured the simple and various photothermal effects by replacing the suspension with a PZT. We used a PPLN crystal to induce the photothermal effect with a high incident power of 600\,mW (measurement error: $\pm7$\%).\par
Figures\,\ref{fig:PT-spectrumetc}(a) and\,\ref{fig:PT-spectrumetc}(b) show the transmitted power $P_\mathrm{t}$ measured using the cavity scan. Figure\,\ref{fig:PT-spectrumetc}(a) shows the spectrum obtained as the mirror moves in the direction of increasing cavity length, which appears to be narrower at the half-maximum width than in the case without the photothermal effect. Figure\,\ref{fig:PT-spectrumetc}(b) shows the spectrum with the mirror moving in the opposite direction, in which case it takes an extremely long time to reach the resonant state. These trends are in agreement with the simulation results shown in Fig.\,\ref{fig:PT-outline}(b), where the effective cavity length increases as the intracavity power increases, indicating that the coefficient of the photothermal effect $\alpha$ is positive.\par
As shown in Sec.\,\ref{sec:PT-TFofcavity}, the modification of the effective cavity length by the photothermal effect causes the cavity to exhibit a frequency response. Using an actuator with a resonant frequency that is sufficiently higher than that of the optical spring, such as a PZT, we can directly measure the optical response of the cavity $H_\mathrm{th}$. Figure\,\ref{fig:PT-spectrumetc}(c) shows the optical response of the cavity. Because the gain of $H_\mathrm{th}$ varies with the PD and electrical driver, only the phase measurement results are shown here. At the initial phase of this measurement, the thermal lens effect of the PPLN caused a severe mode mismatch, so we repositioned the PPLN crystal to pass a 300-$\mu$m beam waist. The crystal clipped the beam and the finesse was reduced to approximately $\mathcal{F}=70\pm10$. Even under these conditions, we observed a noticeable change in the optical response of the cavity because we achieved the condition $\omega_\mathrm{th}>\gamma_\mathrm{th}$. As shown in Eq.\,(\ref{eq:PT-cavityTF}), $H_\mathrm{th}$ is the phase-lead compensation when $\alpha>0$ and $\xi_0>0$. However, when $\alpha>0$ and $\xi_0<0$, the phase becomes approximately $-180$\,degrees in the bandwidth below $\gamma_\mathrm{th}$.\par
The measured results are inconsistent with the theory in the band below approximately 15\,Hz. When the signal was varied slowly, the effective heat capacity may have increased because the region that contributes to the heat outflow became wider. Therefore, the effective $\gamma_\mathrm{th}$ exhibits a frequency response that decreases in the low-frequency band. The measured data becomes consistent with the theory if we use only data above 15\,Hz for fitting. The estimated parameters are $\omega_\mathrm{th}/2\pi=51.7 \pm 4.4$\,Hz and $\gamma_\mathrm{th}/2\pi=12.0 \pm 1.5$\,Hz for $\xi_0=1.00$, and $\omega_\mathrm{th}/2\pi=-79.3 \pm 20.5$\,Hz and $\gamma_\mathrm{th}/2\pi=8.95 \pm 5.4$\,Hz for $\xi_0=-1.00$. In the case of $\xi_0<0$, the fitting accuracy may have been low because $H_\mathrm{th}$ does not significantly change even if the parameters are changed significantly.\par
When the cavity length changes owing to the photothermal effect, we can self-lock the cavity without the use of any feedback control mechanism. In particular, because our experimental system was equipped with a crystal heater for phase matching, we were able to control the cavity near the resonance point without the use of an actuator. Figure\,\ref{fig:PT-spectrumetc}(d) shows the transmitted power and temperature during the self-locking process. We installed the PPLN crystal to pass a 300-$\mu$m beam waist once again for this experiment. The crystal was kept at room temperature and the cavity was out of resonance at $0\sim60$\,s. We switched on the heater at $60\sim79$\,s, and the effective cavity length increased by several $\mu$m because of thermal expansion as the crystal temperature increased. After $79$\,s, heat slowly flowed out of the entire crystal and the effective cavity length decreased. However, after $130$\,s, the intracavity power became more robust, and the heat inflow due to carrier light absorption began. When these two factors were balanced, the crystal temperature and intracavity power stabilized and the self-locking of the cavity was achieved. 
	
\subsection{Measurement of the optomechanical response function with the photothermal effect}\label{sec:PT-OSwPT}

\begin{figure}[htbp]
	\centering
	\includegraphics[width=\hsize]{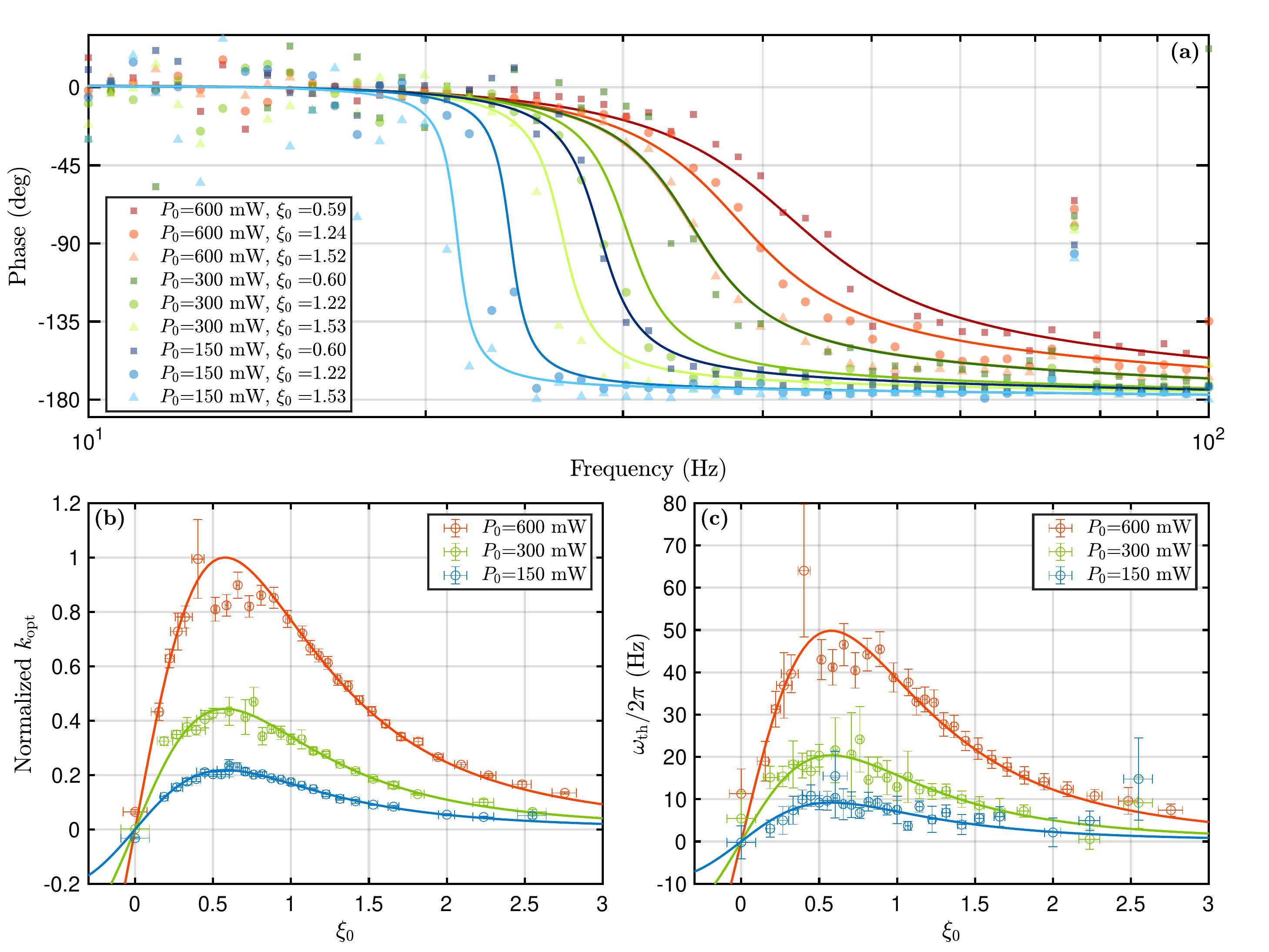}
	\caption{Optomechanical response function with the photothermal effect and parameter estimation. The input power is 600, 300, and 150\,mW as indicated in red, green, and blue, respectively. \textbf{(a)} Representative phase measurement results of the optomechanical response function. The normalized cavity detuning is $\xi_0\sim0.6$ for the squares, $\xi_0\sim1.2$ for the circles, and $\xi_0\sim1.5$ for the triangles. The solid line corresponding to each color represents the fitting results with the optical spring constant $k_\mathrm{opt}$ and photothermal absorption rate $\omega_\mathrm{th}$ as parameters. \textbf{(b)(c)} Estimation results for $k_\mathrm{opt}$ and $\omega_\mathrm{th}$. The circles with error bars correspond to the estimation results achieved by fitting the response function, as shown in (a). The error bar on the vertical axis represents the standard error obtained by fitting, and the error bar on the horizontal axis represents the setting error of $\xi_0$ estimated from the fluctuation of transmitted power. The solid line of each color corresponds to the fitting results obtained using these estimation results with the maximum values of $k_\mathrm{opt}$ and $\omega_\mathrm{th}$ as parameters. The vertical axis in (b) is normalized by the maximum value of $k_\mathrm{opt}$ that was estimated for $P_0$=600\,mW.}
	\label{fig:PT-OSmeasured}
\end{figure}

From Eq.\,(\ref{eq:PT-Koptth}), the imaginary component of the optical spring constant reaches its maximum value when $\omega_\mathrm{th}\sim\gamma_\mathrm{th}$, and when $\omega_\mathrm{th}$ is increased further, the decrease in the real component of the optical spring constant cannot be ignored. The magnitude of $\omega_\mathrm{th}$ depends on the input power $P_0$ and normalized cavity detuning $\xi_0$, but when $P_0=600$\,mW and $\xi_0=1.00$, $\omega_\mathrm{th}$ of the PPLN crystal is approximately four times larger than $\gamma_\mathrm{th}$, and the complex optical spring constant is not adequately large for both the real and imaginary components. However, because the PPKTP crystal achieved approximately $\omega_\mathrm{th}\sim\gamma_\mathrm{th}$ under these parameters and the thermal lens effect was negligible, we measured the optical spring constants using this crystal.\par
Initially, the characteristics of the mechanical suspensions were examined. The angular frequency of the mechanical resonance was roughly estimated to be $\Omega_\mathrm{m}/2\pi=14.2\pm0.1$\,Hz through the measurement of the oscillation magnitude with the signal applied to the coil. The mechanical Q factor was estimated to be $Q_\mathrm{m}=m\Omega_\mathrm{m}/\Gamma_\mathrm{m}=193\pm3$ through the ringdown measurement using the shadow sensing method. The contribution of the optical spring to the damping loss angle can be calculated as $\Gamma_\mathrm{opt}/(m\Omega_\mathrm{opt})\simeq-\Omega_\mathrm{opt}/\gamma\sim-10^{-5}$, where $\Omega_\mathrm{opt}=\sqrt{k_\mathrm{opt}/m }$ is the resonant angular frequency of the optical spring. Without the photothermal effect, the mechanical damping was dominant and the optical damping was negligible.\par
Because $\gamma_\mathrm{th}$ is constant regardless of the input power and cavity detuning, we estimated it from $H_\mathrm{th}$ measurements using a PZT. We measured $H_\mathrm{th}$ for various $P_0$ and $\xi_0$ values but could not estimate the parameters when the condition $\omega_\mathrm{th}\gtrsim\gamma_\mathrm{th}$ was not satisfied. We performed multiple measurements with parameters for which $\omega_\mathrm{th}$ was sufficiently large, and by taking their weighted average, we estimated $\gamma_\mathrm{th}=30.0\pm0.3$\,Hz.\par
Based on the estimated results for these parameters, we investigated the influence of the photothermal effect on the optical spring. The setup shown in Fig.\,\ref{fig:PT-setup}(a) was used to measure the combined spring constant owing to the optical spring and mechanical suspension system. The optomechanical response function is $\chi_\mathrm{eff}H_\mathrm{th}$, from the force $\delta F_\mathrm{ext}(\Omega)$ applied to the test mass for the effective cavity length change $\delta x(\Omega)$. We set $P_0$ to three patterns of 600, 300, and 150\,mW, and varied $\xi_0$ finely in the range of approximately $0\sim3$. When 600\,mW was injected, and the cavity was in the resonance state and the most intense second harmonic was generated. Even then, the intracavity loss estimated from the reflected power measurement was approximately 0.074 times the input coupler loss, and therefore the condition of overcoupling was satisfied. \par
Figure\,\ref{fig:PT-OSmeasured}(a) shows a representative sample of the phase measurement results of the optomechanical response function. We performed the fitting using the optical spring constants $k_\mathrm{opt}$ and $\omega_\mathrm{th}$ as the parameters. As we have shown in Eqs.\,(\ref{eq:PT-TFparameter}) and\,(\ref{eq:PT-kopt0}), $k_\mathrm{opt}$ and $\omega_\mathrm{th}$ exhibit the same functional dependence on $P_0$ and $\xi_0$. These parameters are maximized when $\xi_0=1/\sqrt{3}\sim0.58$. Here, we show the measured data for approximately $\xi_0\sim0.6,\ 1.2,\ 1.5$. Larger $\xi_0$ values correspond to smaller $k_\mathrm{opt}$ and $\omega_\mathrm{th}$ values.\par
The effect of $k_\mathrm{opt}$ on the optomechanical response function appears at the resonance frequency, which is the frequency at which the phase is approximately $-90$\,degrees. It can be seen that a larger $P_0$ and smaller $\xi_0$ result in a higher resonant frequency. The effect of $\omega_\mathrm{th}$ appears mainly in the optical damping and phase-lead phenomena. Optical damping is caused by converting the real component of the complex optical spring constant into an imaginary component through the photothermal effect. Even if optical damping in the absence of the photothermal effect is negligible, this conversion process can significantly change the damping constant of the entire optomechanical system. The measurement results show that the phase inversion of the optomechanical response function was more gradual than it would in the absence of the photothermal effect. In addition, a change in the optical response of the cavity due to the photothermal effect caused a phase lead. The measured phase is led by more than $-180$\,degrees in a band higher than the resonant frequency.\par
If $P_0$ and $\xi_0$ are chosen to have an identical $k_\mathrm{opt}$ for a certain parameter set, the photothermal absorption rate $\omega_\mathrm{th}$, and thus the response function as well, become identical. For example, although $P_0$ and $\xi_0$ were different for the bright red ($P_0=600$\,mW, $\xi_0=1.52$) and dark green ($P_0=300$\,mW, $\xi_0=0.60$) curves in Fig.\,\ref{fig:PT-OSmeasured}(a), the response functions almost perfectly overlapped, reflecting the fact that $k_\mathrm{opt}$ and $\omega_\mathrm{th}$ were estimated to be nearly identical.\par
The estimation results for $k_\mathrm{opt}$ and $\omega_\mathrm{th}$ are shown in Figs.\,\ref{fig:PT-OSmeasured}(b) and\,\ref{fig:PT-OSmeasured}(c). The circles with error bars represent the estimated values of $k_\mathrm{opt}$ and $\omega_\mathrm{th}$ obtained using the same fitting method as in Fig.\,\ref{fig:PT-OSmeasured}(a). In these measurements, we varied $\xi_0$ such that the ranges of the transmitted power variations were approximately equal. However, we excluded the measurement results where the parameters were not identifiable in the fitting program, and the variance was estimated to be zero. The solid lines show the fitting results with the inverse of the variance of the estimates as weights. We estimated the maximum values of $k_\mathrm{opt}$ and $\omega_\mathrm{th}$ for each input power. The estimated maximum $k_\mathrm{opt}$ for $P_0=600$\,mW corresponds to $56.1\pm0.5$\,Hz in terms of the resonant frequency of the optical spring.\par

\begin{figure}[htbp]
	\centering
	\includegraphics[width=0.5\hsize]{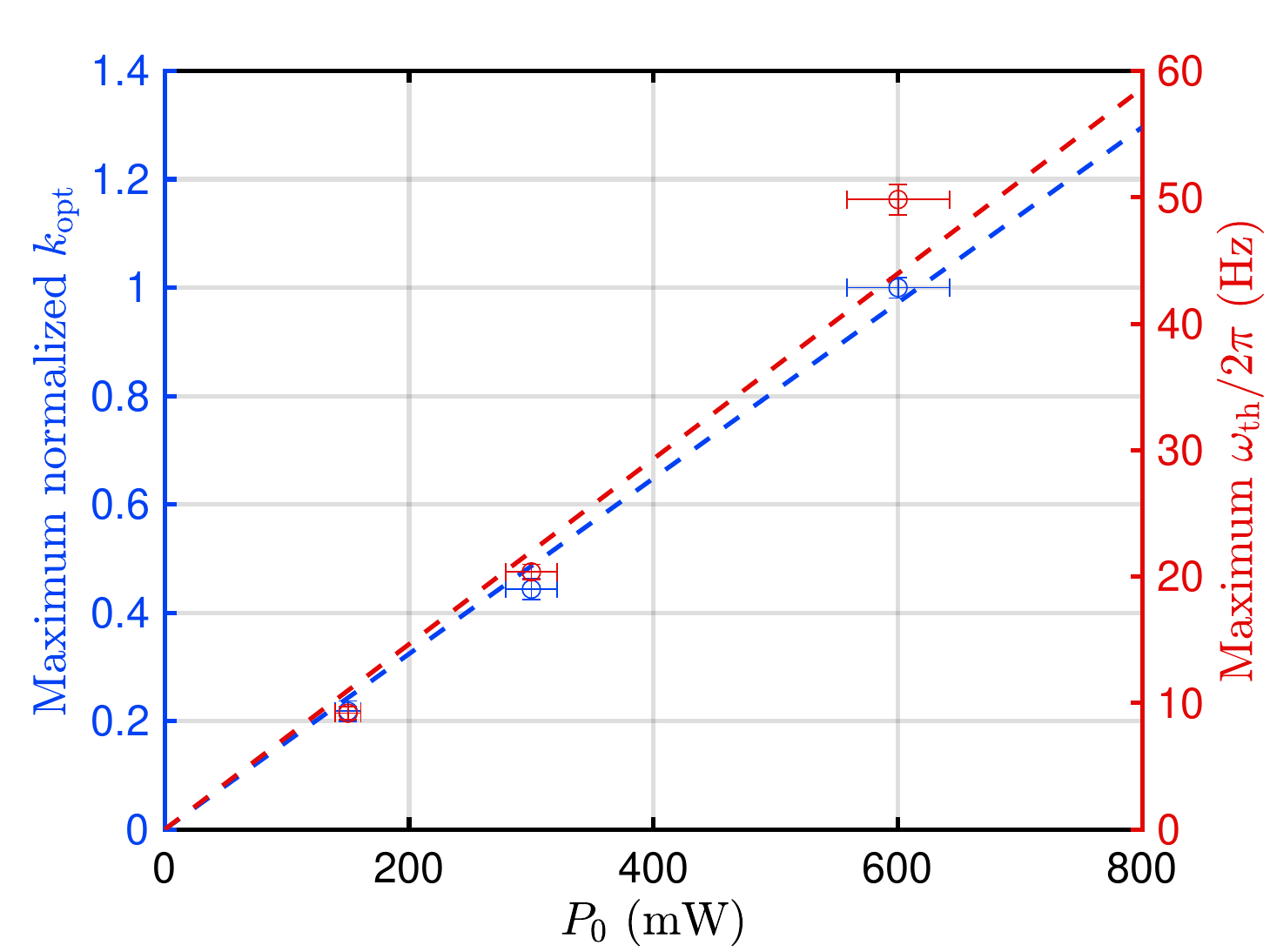}
	\caption{Estimation results for the maximum value of the optical spring constant $k_\mathrm{opt}$ and the photothermal absorption rate $\omega_\mathrm{th}$. The blue circles with error bars show the estimated maximum value of the normalized $k_\mathrm{opt}$, and the red circles with error bars show the estimated maximum value of $\omega_\mathrm{th}$. The error bars on the vertical axis represent the standard error were obtained by fitting, and the error bars on the horizontal axis represent the measurement error of $P_0$. The dotted lines corresponding to each color result from the weighted fitting of these data with a linear function with an intercept of zero.}
	\label{fig:PT-maxmumkomega}
\end{figure}

Figure\,\ref{fig:PT-maxmumkomega} shows the maximum values of $k_\mathrm{opt}$ and $\omega_\mathrm{th}$ estimated from the fitting results. The circles with error bars represent estimates of the respective maximum values, with blue corresponding to $k_\mathrm{opt}$ and red corresponding to $\omega_\mathrm{th}$. The dotted lines show the results of further fitting these estimates with the inverse of their variances as weights, and both are shown to be approximately linear functions with an intercept of zero.

\section{Discussion}
We compare our measurement results with those of a previous experiment using fused silica~\cite{Ma:21}. Although the intracavity power in our experiment was less than $1/10$ of the previous experiment, we observed a comparable photothermal absorption rate $\omega_\mathrm{th}$ because of the more significant thermal absorption coefficient and thermal expansion coefficient of the nonlinear optical crystal. Because there was no significant difference in the specific heat and thermal conductivity of the crystals used in the two experiments, the photothermal relaxation rates $\gamma_\mathrm{th}$ were also comparable. An experiment using PPLN caused serious mode mismatch due to the thermal lens effect, which may be related to the significant thermo-optic coefficient of PPLN compared to other crystals~\cite{Ghosh:94,10.1117/12.804722}.\par
We estimated $\gamma_\mathrm{th}$ by measuring the optical response of the cavity $H_\mathrm{th}$ using a fixed mirror with a PZT. If the beam radius $r_0$ is constant in the part that contributes to thermal absorption, $\gamma_\mathrm{th}$ can be determined using the physical property values of the crystal~\cite{PhysRevD.63.082003},
\begin{equation}\label{key}
	\gamma_\mathrm{th}=\dfrac{\kappa_\mathrm{th}}{\rho C_0 r_0^2},
\end{equation}
where $\kappa_\mathrm{th}$ denotes the thermal conductivity, $\rho$ denotes the density, and $C_0$ denotes the specific heat capacity. For the experiment using PPKTP, the beam radius of the waist is 40\,$\mu$m, the crystal length is 10\,mm, the refractive index is 1.7, the thermal conductivity of KTP is $\kappa_\mathrm{th}\sim 2.2$\,$\mathrm{W}/(\mathrm{m}\cdot \mathrm{K})$~\cite{10.1117/12.239654}, $C_0=6.9\times10^2$\,$\mathrm{J}\cdot\mathrm{kg}\cdot\mathrm{K}$, and $\rho=3.0\times 10^3$\,$\mathrm{kg}/\mathrm{m}^3$, from which the average value of $\gamma_\mathrm{th}/2\pi$ is approximately 95\,Hz. This value is approximately 3.2 times higher than that measured in Sec.\,\ref{sec:PT-OSwPT}. This difference may be because of the extra thermal resistance at the junction of the periodic polarization inversion. It should also be noted that the beam radius is sensitive to the position of the curved mirror and crystal; therefore, the estimate of the beam radius used in the calculation may be inaccurate. If the beam radius is accurately measured, this method is sufficiently accurate to be applied to the high-precision estimation of thermal conductivity~\cite{Ma:21}.\par
We estimated $\omega_\mathrm{th}$ in two ways: by measuring the optical response of the cavity $H_\mathrm{th}$ and by measuring the optomechanical response function $\chi_\mathrm{eff}H_\mathrm{th}$. Although $H_\mathrm{th}$ can be measured with a simple experimental system using a fixed mirror, $\omega_\mathrm{th}\gtrsim\gamma_\mathrm{th}$ is required for accurate parameter estimation. Conversely, the measurement of $\chi_\mathrm{eff}H_\mathrm{th}$ using a suspended mirror is a promising parameter estimation method, even when $\omega_\mathrm{th}$ is small, because a minor photothermal effect can induce non-negligible optical damping.\par
To compare the two parameter estimation methods, we calculated the root mean square error (RMSE) for estimating the maximum value of $\omega_\mathrm{th}$. The estimation results obtained using the suspended mirror are presented in Fig.\,\ref{fig:PT-OSmeasured}(c). The RMSEs normalized by the estimated maximum of $\omega_\mathrm{th}$ were 0.0074 for $P_0=600$\,mW of input power, 0.011 for $P_0=300$\,mW, and 0.035 for $P_0=150$\,mW, showing an excellent agreement between the fitting and measurement results. Conversely, when we estimated only $\omega_\mathrm{th}$ using a fixed mirror for similar parameters, the normalized RMSEs were 0.052 for $P_0=$600\,mW, 0.16 for $P_0=$300\,mW, and 0.41 for $P_0=$150\,mW. These are approximately ten times worse than the estimation results obtained using a suspended mirror, which implies that systematic errors in the estimation method using a fixed mirror were not non-negligible when $\omega_\mathrm{th}\lesssim\gamma_\mathrm{th}$. We note that there is also a lower limit for $\omega_\mathrm{th}$ that can be estimated from measurements using a suspended mirror, which is determined by the minimum optical damping that can be measured.\par
There are several other methods for estimating photothermal parameters. One is to measure the temperature decay owing to heat dissipation. $\omega_\mathrm{th}$ can also be estimated from the time required to achieve photothermal self-locking, as introduced in Sec.\,\ref{sec:PT-cavityresponsemeasurement}. However, our investigations shows that neither of these methods worked. We heated the crystal to 50\,$^\circ$C and measured the time required for the crystal to be cooled down to 20\,$^\circ$C, but the estimated parameters were orders of magnitude smaller than the value obtained in the $H_\mathrm{th}$ measurement, probably because of the additional heat capacities of the heater and thermometer. Another option is to estimate the parameters from the spectrum of the cavity scan relying on the qualitative agreement with the simulation results, but this was not achievable because of the poor linearity of the PZT actuator.\par
In these experiments, it was essential to maintain the nonlinear optical crystal in the extreme phase-mismatch condition to avoid nonlinear optical effects. However, a fraction of secondary harmonics was generated and could have affected the measurement in the case with a high input power. In Figs.\,\ref{fig:PT-OSmeasured}(b) and\,\ref{fig:PT-OSmeasured}(c), when the incident light power was $P_0=600$\,mW and the normalized cavity detuning was close to $\xi_0\sim1/\sqrt{3}$, the measurement results somewhat deviated from the fitting function, possibly because of the nonlinear optical effects. Although we have successfully estimated the parameters with reasonable accuracy, the effect of optical loss, which depends on the intracavity power, should be considered when using optical systems that are more susceptible to nonlinear optical effects. \par
The change in the frequency response owing to the photothermal effect is a phenomenon that occurs only in the low-frequency band, and if the measurement frequency band is sufficiently higher than $\omega_\mathrm{th}$, the photothermal effect may be negligible. However, because the optical spring constant $k_\mathrm{opt}$ and photothermal absorption rate $\omega_\mathrm{th}$ exhibit the same functional dependence on the input power, cavity detuning, and finesse, the photothermal effect cannot be avoided by generating a stiff optical spring and increasing the measurement bandwidth. Therefore, unless the physical property values of a nonlinear optical crystal are improved, the photothermal effect must be considered when dealing with intracavity squeezing systems composed of macroscopic and massive test masses. Parameter estimation using the method presented in this paper and predicting the photothermal effect will allow us to correctly discuss the intracavity squeezing effect. \par
Combining intracavity squeezing and the photothermal effect can be a strong tool for manipulating an optical spring. Even when the intracavity power is low, a stiff optical spring can be generated by implementing an intracavity squeezing method~\cite{SOMIYA2016521,KOROBKO20182238}. Moreover, even when only a single carrier is used, the real and imaginary components of the complex optical spring constant can be positive in the frequency band where the photothermal effect is dominant~\cite{PhysRevD.92.062003,Altin2017}. Optical springs have two inherent problems as they are unstable on their own and the intracavity power limits the magnitude of the spring constant. However, combining the photothermal effect and intracavity squeezing can solve these two problems simultaneously. It is also worth noting that the optical spring generated in such a system has high design flexibility. Although the method described in Sec.\,\ref{sec:PT-THEORY} is helpful for more advanced theoretical calculations, such as when implementing an intracavity squeezing technique, a derivation using Hamiltonian notation is provided in \href{https://doi.org/10.6084/m9.figshare.21333057}{Supplement 1} for a more detailed discussion. We conclude that intracavity squeezing can enhance the optical spring constant and photothermal absorption rate by the same factor.

\section{Conclusion}
In this study, we investigated the influence of the photothermal effect on intracavity squeezing systems. As a result, we found that the photothermal effect profoundly influences the optical response of the cavity and susceptibility of the optomechanical oscillator. When dealing with intracavity squeezing systems composed of macroscopic test masses, the influence of photothermal effects must necessarily be considered. Experimentally, the measured susceptibility of the optomechanical oscillator with the photothermal effect agreed with the theoretical model. The resonant frequency of the optical spring in the absence of the photothermal effect was accurately estimated with a standard error of less than 1\%. We also demonstrated that even a minor photothermal effect can be estimated more accurately than via the previously developed method by measuring the optomechanical response function. While the intracavity squeezing method is known to help increasing the optical spring constant, the photothermal effect can play a complimentary role to stabilize the optical spring. An appropriate combination of the two techniques allows us to design a sensitive and stable optomechanical probe for the precise measurement.

\begin{backmatter}
\bmsection{Funding}
Core Research for Evolutional Science and Technology (JPMJCR1873); Japan Society for the Promotion of Science (20J22778); Sumitomo Foundation (200629).
\bmsection{Acknowledgments}
We would like to thank Jerome Degallaix and colleagues at the LMA for providing us a specially coated mirror, and we would like to thank John Winterflood from UWA for designing the double spiral spring.
\bmsection{Disclosures}
The authors declare no conflicts of interests.
\bmsection{Data availability}
Data underlying the results presented in this paper are not publicly available at this time but may be obtained from the authors upon reasonable request.
\bmsection{Supplemental document}
See \href{https://doi.org/10.6084/m9.figshare.21333057}{Supplement 1} for the supporting content. 
\end{backmatter}

\bibliography{PTbib}

\end{document}